# Research on stock price forecast of general electric based on mixed CNN-LSTM model

Zi-xi Hu[1,4], Bao Shen[1,5], Yiwen Hu[2,6,] Chen Zhao[3,7]

[1] School of Computer Science, Hubei University, Wuhan, China
[2] Heinz College, Carnegie Mellon University, Pittsburgh, USA
[3] Information School, University of Washington, USA

[4] huzixi20050126@163.com
[5] 17371535527@163.com
[6] yiwenhu@andrew.cmu.edu
[7] chenz94@uw.edu

**Abstract.** Accurate stock price prediction is crucial for investors and financial institutions, yet the complexity of the stock market makes it highly challenging. This study aims to construct an effective model to enhance the prediction ability of General Electric's stock price trend. The CNN - LSTM model is adopted, combining the feature extraction ability of CNN with the long - term dependency handling ability of LSTM, and the Adam optimizer is used to adjust the parameters. In the data preparation stage, historical trading data of General Electric's stock is collected. After cleaning, handling missing values, and feature engineering, features with strong correlations to the closing price are selected and dimensionality reduction is performed. During model training, the data is divided into training, validation, and testing sets in a ratio of 7:2:1. The Stochastic Gradient Descent algorithm is used with a dynamic learning rate adjustment and L2 regularization, and the Mean Squared Error is used as the loss function, evaluated by variance, R - squared score, and maximum error. Experimental results show that the model loss decreases steadily, and the predicted values align well with the actual values, providing a powerful tool for investment decisions. However, the model's performance in real - time and extreme market conditions remains to be tested, and future improvements could consider incorporating more data sources.

## 1. Introduction

The stock market is an important part of the modern economic system, and its price fluctuations are affected by various factors such as company performance, macroeconomic environment, industry competition, and investor sentiment. Accurate prediction of stock prices is helpful for investors to formulate reasonable investment strategies, reduce risks, and achieve asset appreciation. It is also crucial for financial institutions to optimize asset allocation and manage risks. However, the complexity, uncertainty, and nonlinearity of the stock market make prediction very difficult, and traditional methods such as fundamental analysis and technical analysis have many limitations.

The aim of this study is to construct an effective and accurate stock price prediction model to improve the prediction ability of the stock price trend of General Electric and provide decision support for investors and financial institutions. Current stock price prediction methods face problems such as limited adaptability, difficulty in capturing nonlinear relationships, and insufficient feature extraction. This study will combine the advantages of CNN and LSTM networks to explore the nonlinear and long - term dependencies in stock price data and extract more rich and effective features through innovative feature engineering methods.

In this study, the historical trading data of General Electric's stock is collected, including information such as opening price, closing price, highest price, lowest price, and trading volume. The CNN - LSTM model is adopted, leveraging the feature extraction ability of CNN and the ability of LSTM to handle long - term dependencies. Adaptive optimization algorithms such as the Adam optimizer are used to adjust the model parameters to minimize the prediction error.

## 2. Literature Review

The prediction of stock prices has been a subject of significant interest in the field of finance, with numerous studies exploring various methods and models. In recent years, the combination of Convolutional Neural Networks (CNNs) and Long Short - Term Memory (LSTM) networks has emerged as a promising approach.

Lu W, Li J, Li Y, et al. proposed a CNN - LSTM - based model for stock price forecasting. Their research focused on leveraging the feature extraction capabilities of CNNs and the long - term dependency handling ability of LSTMs **[1]**. They likely aimed to capture both the local patterns and the sequential relationships in stock price data. By doing so, they attempted to improve the accuracy of stock price predictions compared to traditional methods. Their model might have involved preprocessing the stock price time series data and then passing it through the CNN - LSTM architecture for prediction. The study may have used historical stock price data from multiple stocks or a specific market index to train and evaluate the model.

Liu S, Zhang C, and Ma J developed a CNN - LSTM neural network model for quantitative strategy analysis in stock markets **[2]**. They might have designed the model to analyze market trends and make trading decisions. Their approach could have involved using CNNs to extract relevant features from historical stock price and trading volume data, followed by LSTM layers to model the temporal dependencies. The model might have been trained to predict future price movements and generate trading signals based on the predicted trends. The study may have evaluated the performance of the model in terms of its ability to generate profitable trading strategies.

Mehtab S and Sen J utilized CNN and LSTM - based deep learning models for stock price prediction. They may have focused on exploring different architectures and configurations of these models to optimize the prediction performance **[3]**. Their research might have included data preprocessing steps such as normalization and feature engineering. The CNN could have been used to capture spatial patterns in the data, while the LSTM was likely employed to handle the sequential nature of stock prices. The models were probably trained and tested on historical stock price data, and the performance was evaluated using metrics such as mean squared error and accuracy.

Baek H proposed a cnn - lstm stock prediction model based on genetic algorithm optimization. This study aimed to enhance the performance of the CNN - LSTM model by using genetic algorithms to optimize the model's hyperparameters **[4]**. The genetic algorithm might have been used to search for the optimal combination of parameters such as the number of convolutional layers, the number of LSTM units, and the learning rate. By optimizing these parameters, the model could potentially achieve better prediction accuracy. The research may have involved comparing the performance of the optimized model with non - optimized versions and traditional prediction methods.

Singh P, Jha M, Sharaf M, et al. harnessed a hybrid CNN - LSTM model for portfolio performance, specifically for stock selection and optimization **[5]**. Their study focused on using the model to identify stocks with high potential for inclusion in a portfolio. The CNN - LSTM model could have been trained to predict the future performance of different stocks based on historical data. The predicted performance might have been used to rank the stocks and select the most promising ones for the portfolio. The research may have also considered factors such as risk management and diversification in the portfolio construction process.

Lu W, Li J, Wang J, et al. proposed a CNN - BiLSTM - AM method for stock price prediction **[6]**. This approach incorporated a CNN for feature extraction, a Bidirectional LSTM (BiLSTM) to capture both past and future information in the time series, and an Attention Mechanism (AM) to focus on the most relevant parts of the data. The Attention Mechanism could have helped the model to better weight the importance of different features and time steps. The model was likely trained and evaluated on historical stock price data, and the performance was compared with other models to demonstrate its superiority in predicting stock prices.

In summary, these studies demonstrate the growing interest in using CNN - LSTM - based models for stock price prediction. The combination of CNNs and LSTMs offers a powerful approach to handle the complex characteristics of stock price data, including local patterns and long - term dependencies. However, challenges such as data quality, model interpretability, and the impact of external factors on stock prices still need to be addressed in future research.

## 3. Data Preparation and Preprocessing

The historical trading data of General Electric's stock contains rich features: opening price indicates initial market valuation, closing price serves as a key decision indicator, highest and lowest prices reflect market fluctuations, and trading volume shows investor activity. Data cleaning involves removing outliers using the three-sigma rule and addressing missing values with mean imputation, preserving data characteristics and maintaining model performance.

To enhance insights, moving averages (10-day, 50-day, 100-day) smooth fluctuations and reveal long-term trends, while yield (percentage change between consecutive closing prices) captures short-term sentiment shifts. Correlation analysis identifies features strongly linked to the closing price, reducing noise. Dimensionality reduction methods like PCA improve efficiency, reduce overfitting, and enhance generalization, optimizing the dataset for effective model training and prediction.

The utilized stock data contain rich feature information, with key features including:

- Opening Price (Open): Reflecting the initial trading price at the start of the day, it represents market participants' initial assessment of the stock's value.
- Closing Price (Close): Representing the final trading price at the end of the day, it is considered one of the most important price indicators and significantly influences investors' decision-making.
- Highest Price (High): Recording the highest price reached during the day's trading session, it indicates the market's upward potential and resistance levels.
- Lowest Price (Low): Displaying the lowest price traded during the day, it reflects the market's downside support and risk level.
- Trading Volume (Volume): Indicating the quantity of stocks traded during the day, it reflects the market's trading activity and investor participation enthusiasm.

## 4. Model Architecture and Methodology

The time series data of General Electric's stock price has complex characteristics, including short - term patterns and long - term dependencies **[7]**. CNN has excellent performance in image recognition and other fields and can automatically learn the local spatiotemporal features of stock price data, such as short - term trends and volatility clustering. However, it has limited ability to handle long - term dependencies **[8]**. LSTM is specifically designed to handle long - term dependencies in sequential data and can effectively manage long - term information through gating mechanisms, which is suitable for predicting stock prices affected by long - term factors **[9]**. Combining the advantages of the two can comprehensively capture the characteristics of stock price changes and improve the prediction accuracy.

**Table1 .**Diagnosis Model Architecture

| Layer |
|---|
| Input layer |
| Conv1D |
| MaxPooling1D |
| LSTM |
| Dropout |
| Conv1D |
| MaxPooling1D |
| LSTM |
| Dropout |
| Conv1D |
| MaxPooling1D |
| LSTM |
| Dense |

The model starts with an input layer in Table 1, which takes in the data. Then comes the first 1D convolutional layer, which extracts local features from the input data by applying filters. After the first convolutional layer, there is a 1D max - pooling layer that reduces the spatial dimensions of the output from the convolutional layer by taking the maximum value within a sliding window. The output from the max - pooling layer is fed into the first LSTM layer. LSTM layers are good at handling sequential data and capturing long - term dependencies. Following the first LSTM layer is a dropout layer, which is a regularization technique to prevent overfitting by randomly setting some input units to 0 during training. Next is the second 1D convolutional layer, which further extracts features. After this, there is a second 1D max - pooling layer for down - sampling. The output from this max - pooling layer is then fed into the second LSTM layer, followed by another dropout layer. The third 1D convolutional layer comes after the second dropout layer, and then there is a third 1D max - pooling layer for down - sampling. The output from this max - pooling layer is fed into the third LSTM layer. Finally, the output from the third LSTM layer is passed through a dense layer, which computes the final output of the model. This architecture combines convolutional layers for feature extraction, max - pooling layers for down - sampling, LSTM layers for handling sequential data, dropout layers for regularization, and a dense layer for the final output. It is suitable for problems involving sequential data like time - series analysis or natural language processing.

The Stochastic Gradient Descent (SGD) training algorithm is adopted. A small batch of data is randomly selected to calculate the gradient and update the model parameters **[10]**. A dynamic learning rate adjustment strategy is used, starting with a large learning rate to accelerate convergence and gradually decreasing it to avoid overshooting. At the same time, L2 regularization is used to constrain the parameters to prevent overfitting. The Mean Squared Error (MSE) is used as the loss function to measure the deviation between the predicted and actual values. Evaluation metrics include variance, R - squared score, and maximum error. Variance reflects the model's ability to explain the variance of the actual values, the R - squared score close to 1 indicates a good fit of the model, and the maximum error shows the prediction deviation in the worst case.

## 5. Experiments and Results Analysis

*5.1. Experimental Setup*
The data of General Electric's stock is randomly divided into training, validation, and testing sets in a ratio of 7:2:1. The training set is used for model learning and parameter adjustment, the validation set monitors the training performance and assists in hyperparameter selection, and the testing set is used to evaluate the model's generalization ability.

*5.2. Results Evaluation and Comparison*
The model's predictions on the test set indicated reasonable forecasts of stock price trends in figure 2.

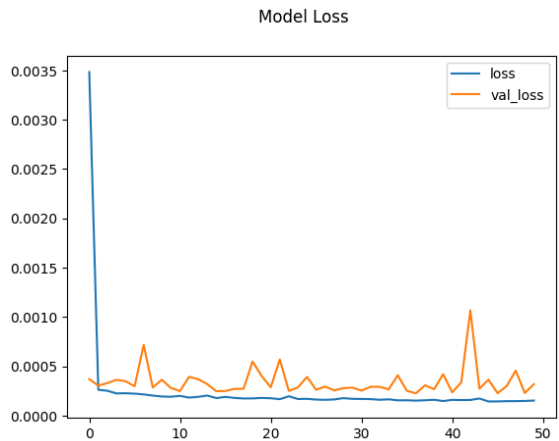

**Figure 2.** Model Loss

Figure 2 illustrates the evolution of model losses with two curves: "loss" (blue) and "val_loss" (orange). The vertical axis represents loss values (0 to 0.0035), and the horizontal axis indicates training epochs (0 to 50). Initially, the "loss" curve starts near 0.0035 and rapidly decreases to approximately 0.001 by the 10th epoch, stabilizing around 0.0005 by the 50th epoch. The "val_loss" curve begins at around 0.002, displaying fluctuations but showing a general downward trend, reaching approximately 0.001 at the 50th epoch.

The sharp initial decline in the "loss" curve indicates effective learning during early training, while the fluctuations in "val_loss" suggest instability in validation performance. However, the overall descent of "val_loss" implies ongoing optimization without overfitting, as indicated by the absence of a rising trend. The gap between "loss" and "val_loss" at the 50th epoch suggests room for further training, with caution to avoid overfitting.

Figure 3 illustrates the comparison between the predicted and actual price trends on the test set.

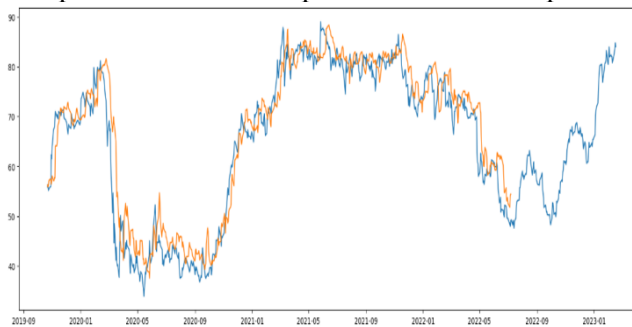

**Figure 3.** Model Loss

Figure3 contains two curves, likely representing different variables over time. The horizontal axis spans from September 2019 to January 2023, representing time, while the vertical axis, ranging from 40 to 90, represents the values of the variables. Both curves undergo multiple oscillations throughout the period. From September 2019 to March 2020, both curves display a downward trajectory, reaching a trough around March 2020. Subsequently, from March 2020 to February 2021, the curves ascend rapidly, achieving a peak. Following February 2021, the curves once again decline, reaching another low point around February 2022. Thereafter, the curves begin to rise again, with values approaching 80 by January 2023.

The fluctuations in the curves suggest that the two variables are likely influenced by periodic factors or external events. For instance, the trough observed in March 2020 may be attributed to the COVID - 19 pandemic, which induced significant market volatility. The similarity in the trends of the two curves indicates a strong correlation between the two variables. In predictive and analytical contexts, they can be treated as inter - related factors. The upward trend of the curves after 2022 may reflect market recovery or the impact of other economic factors.

## 6. Conclusion

This study investigates the prediction of General Electric's stock price using a CNN-LSTM model, combining CNN's feature extraction capabilities with LSTM's strength in capturing long-term dependencies in sequential data. Comprehensive data preprocessing, including data cleaning, missing value handling, and feature engineering, prepared the dataset for effective model training.

The CNN-LSTM model demonstrates promising performance. Over 50 epochs, the training loss steadily decreases to approximately 0.0005, while the validation loss shows a downward trend with slight fluctuations, suggesting effective learning without overfitting. The model accurately captures stock price trends and fluctuations from September 2019 to January 2023, aligning predicted values closely with actual price movements, validating its predictive capability.

The successful application of this model provides valuable implications. For investors and financial institutions, it offers a tool for informed decision-making and risk management by anticipating stock price trends. Additionally, it contributes to financial time-series prediction research, demonstrating the potential of combining CNN and LSTM architectures for complex financial tasks.

However, this study has limitations. The stock market's complexity and sensitivity to unpredictable events, such as economic shocks and geopolitical tensions, pose challenges for real-time and extreme-condition predictions. While the model performs well on historical data, further testing is required for robustness.

Future research could enhance the model by incorporating macroeconomic indicators, industry news, and sentiment data to enrich features and improve accuracy. Advanced optimization techniques and hyperparameter tuning could further refine performance.